\documentclass[preprint,showpacs,preprintnumbers,amsmath,amssymb,nofootinbib]{revtex4}

\usepackage{amsthm}

{Corollary}[section]
{Proposition}[section]

\usepackage{mathrsfs}
\usepackage{etex}
\usepackage{amssymb,amsthm,amscd,amsbsy,array}
\usepackage{bm}
\usepackage{amsmath,dsfont}
\usepackage{soul} 
\usepackage{graphics,graphicx,xcolor}

\usepackage[colorlinks=true, pdfstartview=FitV, linkcolor=blue, citecolor=blue, urlcolor=blue]{hyperref} 

\newcommand{\gb}{\quad\colorbox{green}}

\newenvironment{redtext}{\color{red}}
{\ignorespacesafterend}
\newenvironment{bluetext}{\color{blue}}{\ignorespacesafterend}
\newenvironment{greentext}{\color{green}}{\ignorespacesafterend}
\newenvironment{magentatext}{\color{magenta}}{\ignorespacesafterend}
\newenvironment{cyantext}{\color{cyan}}{\ignorespacesafterend}
\newenvironment{orangetext}{\color{orange}}
{\ignorespacesafterend}

\newcommand{\bmagenta}{\begin{magentatext}}
\newcommand{\emagenta}{\end{magentatext}}
\newcommand{\bcyan}{\begin{cyantext}}
\newcommand{\ecyan}{\end{cyantext}}
\newcommand{\bblue}{\begin{bluetext}}
\newcommand{\eblue}{\end{bluetext}}
\newcommand{\bred}{\begin{redtext}}
\newcommand{\ered}{\end{redtext}}

\newcommand{\bgreen}{\begin{greentext}}
\newcommand{\egreen}{\end{greentext}}
\newcommand{\borange}{\begin{orangetext}}
\newcommand{\eorange}{\end{orangetext}}

\numberwithin{equation}{section}

\let\ssection=\section
\renewcommand{\section}{\setcounter{equation}{0}\ssection}
\newcommand{\beq}{\begin{equation}}
\newcommand{\eeq}{\end{equation}}
\newcommand{\bec}{\begin{center}}
\newcommand{\ec}{\end{center}}

\newcommand{\bb}{{\mathbf{b}}}

\newcommand{\bbeta}{\boldsymbol{\beta}}

\newcommand{\bx}{{\bm{x}}}
\newcommand{\bX}{{\bm{X}}}

\newcommand{\bc}{{\mathbf{c}}}

\def\aand{{\quad\text{\small and}\quad}}
\def\where{{\quad\text{\small where}\quad}}

\newcommand{\bd}{{\mathbf{D}}}

\newcommand{\cE}{{\mathcal{E}}}

\newcommand{\cL}{{\mathscr{L}}}
\newcommand{\cS}{{\mathscr{S}}}

\newcommand{\bR}{{\bf R}}

\newcommand{\bU}{{\bf U}}

\newcommand{\bv}{{\bf v}}

\newcommand{\bW}{{\bf W}}

\def\smallover\#1/\#2{\hbox{$\textstyle\frac{\#1}{\#2}$}} %

\def\bv{{\bm{v}}}

\def\parag{\hfil\break} 
\def\kikezd{\parag\underbar}

\def\bequ{\begin{enumerate}}
\def\eenu{\end{enumerate}}
\def\bitem{\begin{itemize}}
\def\eitem{\end{itemize}}

\def\beq{\begin{equation}}
\def\eeq{\end{equation}}
\def\beqa{\begin{eqnarray}}
\def\eeqa{\end{eqnarray}}

\def\barray{\left(\begin{array}}
\def\earray{\end{array}\right)}
\def\barraynb{\begin{array}}
\def\earraynb{\end{array}}

\def\IR{{\mathbb{R}}} 

\def\?{{\,\gb{\fbox{\texttt{??}}\;}}\,}

\def\Rarrow{{\quad\Rightarrow\quad}}

\def\benu{\begin{enumerate}}
\def\eenu{\end{enumerate}}
\def\bitem{\begin{itemize}}
\def\eitem{\end{itemize}}

\newcommand{\const}{\mathop{\rm const.}\nolimits}
\newcommand{\half }{\smallover{1}/{2}}

\def\smallover#1/#2{\hbox{$\textstyle\frac{#1}{#2}$}} %
\def\smallcirc{{\raise 0.5pt \hbox{$\scriptstyle\circ$}}}
\def\cabove(#1){\stackrel{\smallcirc}{#1}}
\def\ccabove(#1){\,\stackrel{\smallcirc\smallcirc}{#1}\,}
\def\cccabove(#1){\stackrel{\,\smallcirc\smallcirc\smallcirc}{#1}\,}
\def\2{{\smallover1/2}}

\def\boxit#1{
\vbox{\hrule\hbox{\vrule\kern4pt
\vbox{\kern5pt#1\kern5pt}\kern4pt\vrule}\hrule}
} 

\newcommand{\medbox}[1]{\fbox{%
\rule[-10pt]{0pt}{25pt}$\;\;\displaystyle{#1}\;\;$}%
}

\let\ssection=\section
\renewcommand{\section}
{\setcounter{equation}{0}\ssection}

\def\besub{\begin{subequations}}
\def\esub{\end{subequations}}

\begin{document}


\title{MultiCarroll dynamics}
\author{P.-M. Zhang$^{1}$\footnote{corresponding author.  zhangpm5@mail.sysu.edu.cn},
H-X. Zeng$^{1}$\footnote{zenghx53@mail2.sysu.edu.cn} 
and P.~A. Horvathy$^{3}$\footnote{horvathy@lmpt.univ-tours.fr}
}

\affiliation{
${}^1$School of Physics and Astronomy, Sun Yat-sen University, Zhuhai 519082, (China)
\\
${}^{2}$ Institut Denis-Poisson CNRS/UMR 7013 - Universit\'e de Tours - Universit\'e d'Orl\'eans Parc de Grammont, 37200; Tours, (France)
\\
}
\date{\today}


\begin{abstract}
Unlike a single Carroll particle, a multiparticle Carroll
system  \underline{can move} under suitable conditions, as we demonstrate it explicitly for two particles with a momentum-dependent interaction~: the center-of-mass remains fixed, however relative motion \underline{is} possible, confirming  previous statements made by Bergshoeff, Casalbuoni, and their collaborators. 
 Analogous results are obtained for electric dipoles with the roles of the center-of-mass and the relative position interchanged. 
\bigskip

\noindent
Int. J. Theor. Phys. \textbf{63} (2024) no.10, 243
doi:10.1007/s10773-024-05777-7
[arXiv:2306.07002 [gr-qc]].

\bigskip

\noindent{Key words:
Carroll symmetry. (Im)mobility of  Carroll particles.
}
\end{abstract}
\maketitle

\tableofcontents
\goodbreak

\section{Introduction}\label{Intro}

Carroll symmetry, whose characteristic feature is that \emph{Carroll} (or C-) \emph{boosts}  leave the position $\bx$ fixed but shift the ``Carroll time'' (that we denote by $s$ to distinguish it from relativistic or Galilean time),
\beq
\bx \to \bx,
\qquad
s \to s - \bb\cdot\bx\,,
\label{Cboost}
\eeq
has a remarkable history.  
 This unconventional boost-action  
  was obtained by L\'evy-Leblond \cite{Leblond} and (independently) by SenGupta \cite{SenGupta}
   by the $c\to0$  contraction of the Poincar\'e group \cite{BacryLL} --- to be contrasted with the familiar $c\to\infty$ contraction
 introduced by Wigner and In\"on\"u  \cite{Inonu}, which yields the Galilei group. 

 However, apart of occasional studies \cite{Henneaux79,Henneaux82,DGH91,Gibbons:2002tv}, Carroll symmetry has long remained a rather confidential mathematical curiosity 
 \cite{DGH91,Ancille,MarsotJGP} for over 50 years --- before reemerging triumphantly  \cite{Bergshoeff14,Carrollvs}.

How can this strange fate be explained ? 
The main cause may well be that Carroll symmetry, as introduced initially, did not have genuine physical applications --- for the good reason that a \emph{particle with Carroll symmetry does not move} --- a fact recognized from the very beginning that has subsequently become a general wisdom.  

The judgment of ``uselessness'' has changed dramatically, though, when  Carroll symmetry was applied to general relativity, in particular to black holes and gravitational waves \cite{Gibbons:2002tv,Carroll4GW,Donnay19,Donnay22,MarsotHP,HallCarroll}, followed by a huge number of recent papers that we cannot list all.

Another, independent reason for interest in Carroll symmetry comes from  a rather different horizon: it is precisely the ``no motion'' property of the quasiparticles called \emph{fractons} that attracts considerable current attention in condensed matter physics \cite{fractons,Pretkofractons,Seiberg,Gromov:2018nbv,DoshiG,Pena-Benitez:2021ipo,JainJensen,Bidussi}.  Defect has become a virtue~!

The similarity of fractons and  Carroll particles has been noted recently \cite{Bidussi}.
The  ``dipole symmetry'' of condensed matter physicist is indeed a sort of  dual of Carroll symmetry studied in the gravitational context \cite{HallCarroll,FPP,Carroll4GW}.

There is general agreement: a \emph{single} Carroll particle can no move. But what about \emph{multiparticle systems} of Carroll particles~? In  \cite{Bergshoeff14,Casalbuoni} it was argued
\emph{against} their immobility. 
This Note confirms and extends their statements using an approach which is rather different form the one advocated in \cite{Bergshoeff14,Casalbuoni}.
Below we present indeed a simple but explicit counter-example to the ``Carroll-implies-immobility''  dogma. In detail, we demonstrate  that two particles which interact through the momentum-dependent potential
\beq
\label{BGLpot}
V  =  -\half\big(\bv_1-\bv_2\big)^2\,,
\eeq
where the $\bv_a,\, a= 1,2$ represent the {momenta} of particles of unit mass $m_a=1$ (see sec.\ref{Concl}), \emph{do move}. 
Momentum-dependent interactions are rather unconventional  in high-energy physics; however they are widely used in \emph{nuclear physics} \cite{Waak,Korinek,DGGB,Nara}, as we learned it from an expert \cite{Csaba}.
\goodbreak

\section{Immobility and Carroll boost symmetry}

Particles with Carroll symmetry can and in fact have been \cite{Ancille,MarsotJGP} studied systematically by the
``orbit construction'' of Souriau
 \cite{SSD}: a particle in $d$-dimensional space is described by
a $2d + 1$ dimensional ``evolution space'' $\cE = \big\{(\bx, \bv, s)\big\}$ where $\bx$ is the position in configuration space, $\bv$ a tangent vector to the latter, and $s$ is Carrollian time \cite{HallCarroll}. For a translation-invariant system the conserved momentum is proportional to the coordinate $\bv$; therefore we shall refer to $\bv$ simply as ``momentum''.

However, many physicists are not familiar with Souriau's presymplectic geometry \cite{SSD} therefore we find it more convenient to present it in a more conventional, namely Lagrangian framework.
Their relation is explained in \cite{Ury}.

For the one particle Lagrangian $\cL_0 = m \bv \cdot \bx^{\prime}$  \cite{HallCarroll}
the Hamiltonian action
\beq
\cS_0 = \int \!\cL_0 ds = \int\! m\bv\cdot d\bx\,,
\label{CarrHaction}
\eeq
 is invariant under C-boosts \cite{HallCarroll} and also yields the equations of motion,
\beq
\bx^{\prime} = 0 \aand \bv^{\prime} = 0
\Rarrow  \bx=\bx_0 \aand \bv=\bv_0\,.
\eeq
In conclusion, \emph{a  single massive Carroll particle does not move}, consistently with general wisdom \cite{Leblond,SenGupta}.


Let us now turn to multiparticle systems.
First we note that a $d$-dimensional system of
$N$ Carroll particles  with masses $m_a>0$ which do not interact between themselves can be described by a generalization of the 1-particle Carroll action \cite{HallCarroll},
\beq
\cS_0^N= \int\!\cL_0^Nds =\int\left(\sum_{a=1}^N m_a\bv_a\cdot \bx_a^{\prime}\right)ds=
\int\sum_{a=1}^Nm_a\bv_a\cdot d\bx_a\,,
\label{NC0Lag}
\eeq
which gives us  uncoupled equations of motion,
\beq
m_a\bx_a^{\prime}=0\,,
\;
\bv_a^{\prime}= 0
\quad\Rightarrow\quad
\bx_a = \big(\bx_a\big)^0=\const
\;
\bv_a =\big(\bv_a\big)^0=\const
\label{freeNmotion}
\eeq
$a=1,\dots, N$. None of the  particles moves; their (vanishing) velocities, $\bx_a^{\prime}=0$, are unrelated to the constant values of their ``momenta'' $\bv_a$, which  play no dynamical r\^ole.
This is a first hint at that the system reduces to a first-order one.

We then assume that our Carroll particles do interact, namely through an {\rm a priori} arbitrary \emph{position and momentum-dependent} $N$-body potential $V(\bx_a, \bv_a, s)$,
\begin{equation}
\cS = \int \!{\cL}_{int}ds =
\int \Big(\sum_a^N m_a\bv_a \cdot \bx_a' - V\Big) ds = \cS_0 - \int \!V ds \,.
\label{NCVLag}
\end{equation}
The equations of motion which generalize \eqref{freeNmotion} still separate,
\besub
\begin{align}
m_a \bx_a'  = \frac{\partial V}{\partial \bv_a}\,,
\label{xprime}
\\[6pt]
m_a \bv_a'  = - \frac{\partial V}{\partial \bx_a}\,,
\label{vprime}
\end{align}
\label{xvprime}
\esub
$a=1,\dots, N$.
It follows that all of our particles remain fixed, confirming the gneral wisdom,  \emph{unless  the potential depends effectively on the momenta} $\bv_a$, $\frac{\partial V}{\partial \bv_a} \neq 0$, that we shall henceforth assume.

Then our first observation says that when the $\bv$-dependent terms on 
the rhs of  \eqref{xprime}, add up to zero,
\beq
 \sum_a \frac{\partial V}{\partial\bv_a} = 0
 \label{dualNIII}
\eeq
i.e.  when the potential satisfies  a sort of \emph{``momentum dual'' of Newton's 3rd law}, then  the center of mass (CoM),
\beq
{\bX} = \frac{\sum_{a}m_a\bx_a}{M}\,,
\quad 
M=\sum_{a}m_a
\label{CoM}
\eeq
 remains fixed 
\begin{equation}
\medbox{
M \bX^{\prime} =\sum_a \frac{\partial V}{\partial\bv_a} = 0\,.
}
\label{vNIII}
\end{equation}
The second equation, \eqref{xprime}, is in turn irrelevant from the dynamical point of view, because the $\bv_a$ have no bearing on the dynamics which is of the first order. We note however that the ``center of momentum'', a dual to the CoM is fixed by \eqref{vprime},
$
 \sum_a m_a\bv_a = \const
$ 
because of Newton's usual 3rd law. The relative motion can however be non-tivial, as will be seen in Sec.\ref{BGLEx}.

All this follows from the reduction of the 2nd-order Newtonian dynamics to a first-order one,
\eqref{xprime}.

The relation \eqref{vNIII} should be compared with the discussion in \cite{SSD} (pp. 141-142 and pp. 152-153), where Souriau states  ``Galilean relativity + the Maxwell Principle imply Newton's 3rd law''. See also \cite{KohnI,KohnII,ZHAGK,Inzunza}. 

The  boost symmetry is  modified in the Carroll context. The C-boosts \eqref{Cboost} lift to the evolutions space,
\beq
\bx_a \to \;\bx_a\,,\quad \bv_a \to \bv_a+\bb\,,\quad
s \to s -  \bb \cdot \bX\,,
\label{Nboost}
\\[6pt]
\eeq
\!where $\bb\in\IR^d$, and leave the free action \eqref{NC0Lag} invariant up to surface terms,
\beq
 \cL_0 \to \;\cL_0 + \bb \cdot \left(\sum_a {m_a\bx_a}\right)^{\prime}
 =
 \cL_0 + \bb \cdot M \bX\,^{\prime}\,.
\label{freeLchange}
\eeq
The boosts \eqref{Nboost} leave both the CoM \eqref{CoM} and the relative coordinates 
\beq
\bR_{ab}(s)=\bx_a(s)-\bx_b(s)
\label{relcoord}
\eeq
 invariant, ${\bX}\to {\bX}, \;\bR_{ab} \to \bR_{ab}$.
 Then we introduce the relative momenta $\bW_{ab}=\bv_a - \bv_b$ and, generalizing the momentum-dependent Ansatz \eqref{BGLpot} of Bergshoeff et al, limit our investigations to potentials of the form,
\beq
V = V(\bv_a)=\sum_{a > b = 1}^NV_{ab}\big(\bW_{ab}^2\big)\,.
\label{Vvpot}
\eeq
All relative momenta $\bW_{ab}$ and thus the potential \eqref{Vvpot}, 
are also left invariant.

Our assumption ${\partial V}/{\partial \bx_a}=0$ implies that
  $\bv_a=\bv_a^0$. Therefore  $\bW_{ab}(s) = \bW_{ab}^0$
 and thus also the potential are constants of the motion, $V= V(\bv_a^0)=V^0=\const$

The interaction term in \eqref{NCVLag} involves also the Carroll time $s$,
 but  ${\cL}_{int}{ds}$ changes again by a surface term,
\beq
{\cL}_{int}ds = \left(L_{0}-V\right) ds+dK
\where 
K=\left( M+V_{0}\right)  \,\bb\cdot{\bX},
\label{Npichange}
\eeq
allowing us to conclude that the  C-boosts  span a $d$-parameter symmetry.

Remarkably, the  $V^0$ terms cancel in the generated C-boost momentum
${\cL}_{int}\delta s - K$, leaving us with,
\beq
\bd = - M{\bX}\,,
\label{CoMmom}
\eeq
whose conservation plainly implies that ${\bX}$, the
\emph{CoM, can not move} --- it behaves as a single particle.
 However no conclusion can be drawn about the relative motion, $\bR_{ab}(s)=\bx_a(s)-\bx_b(s)$; the two-body examples discussed in the next section shows that $\bR_{ab}$ \emph{can indeed be $s$-dependent}.
\goodbreak

\section{A two-body system of Bergshoeff et al}\label{BGLEx}

A nice illustration for multi-Carroll systems is provided by two  particles with masses $m_1=m_2=1$ (for simplicity) which interact through the momentum-dependent potential $V = - \half\bW^2$ \eqref{BGLpot}, proposed by  Bergshoeff et al \cite{Bergshoeff14}. $V$  plainly satisfies the  condition \eqref{vNIII}. Supplementing our previous notations 
\beq
\bX=\half\big(\bx_1+\bx_2\big),\quad \bR=\bx_1-\bx_2,\quad \bW=\bv_1-\bv_2,
\eeq
 with
$\bU=\half(\bv_1+\bv_2)$, the Hamiltonian action \eqref{NCVLag} can  be decomposed as
\beq
\cS=
\underbrace{\int
2\bU\cdot {\bX}^{\prime}ds
}_{\cS_{X}}\,
\; + \;
\underbrace{\int \half
\bW\cdot {\bR}^{\prime}ds + 
 \int \half\bW^2ds}_{\cS_{R}}\,\,,
\label{GomisHact}
\eeq
whose variational equations,
\beq
\bX^{\prime}= 0\,,
\quad\,
\bU^{\prime}\, = 0\,,
\quad
{\bR}^{\prime}= - 2{\bW},
\quad
\bW^{\prime}= 0,
\label{UWXRvareq}
\eeq
are solved at once by,
\besub
\begin{align}
\bW= &\; \bW^0\,, &\quad \bW^0=\half\big(\bv_1^{0}-\bv_2^{0}\big)=\const
\\
\bX(s)=&\; \bX_0, &\bX_0=\half(\bc_1+\bc_2)=\const
\label{B14CoMfixed}
\end{align}
\label{GomisXR}
\esub
and, even more importantly for us,
\beq\medbox{
{\bR}(s) =  \bR(0)-2\bW^0 \,,
}
\label{B14grelmot}
\eeq
\!supplemented with
$ \bU=\bU^0=\half\big(\bv_1^{0}+\bv_2^{0}\big)=\const $,
 where the $\bv_a^0,\, \bc_a,\ a=1,2$ are integration constants. 
 
\begin{figure}[ht]
\includegraphics[scale=.3]{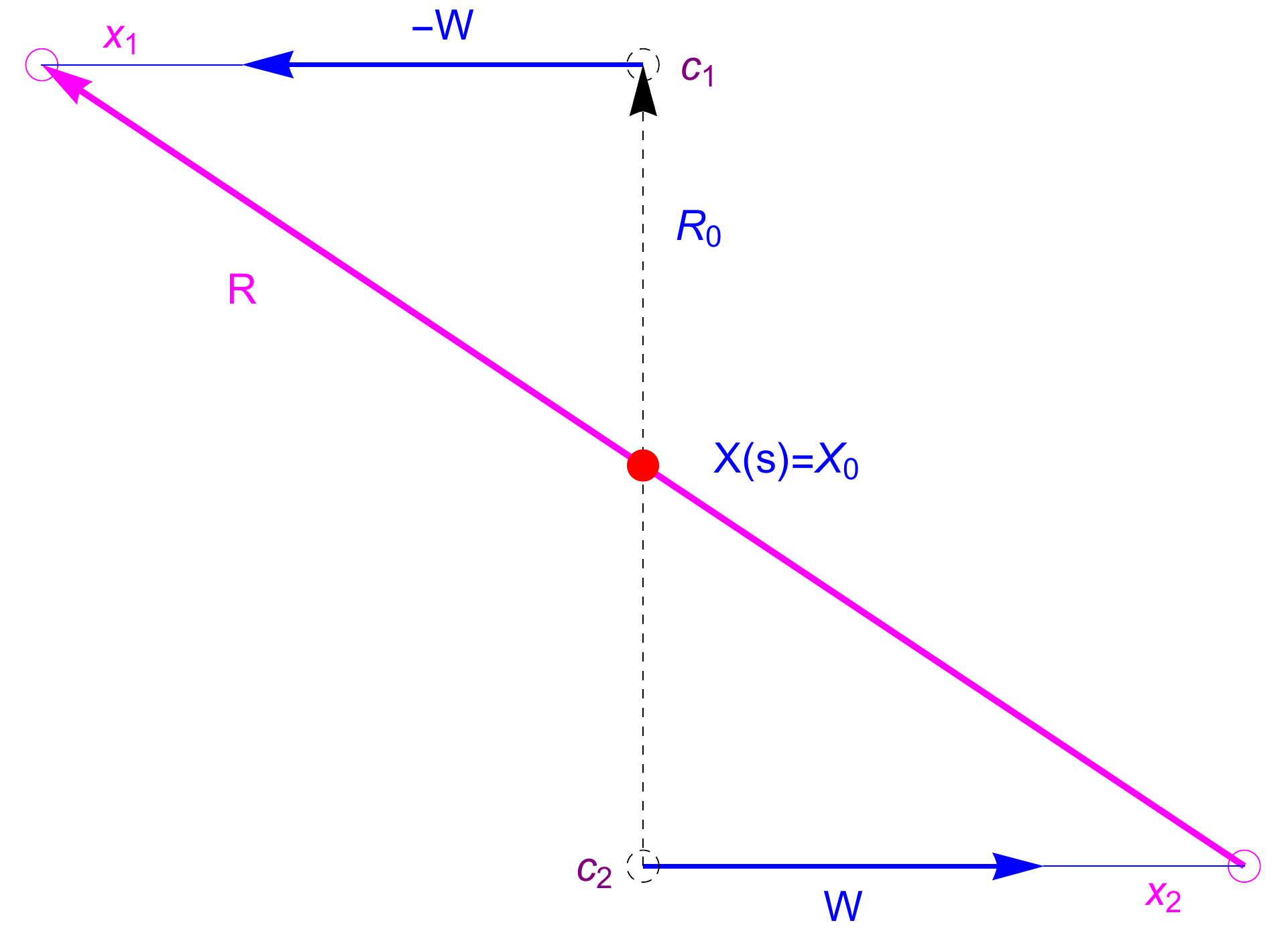}
\hskip16mm
\includegraphics[scale=.27]{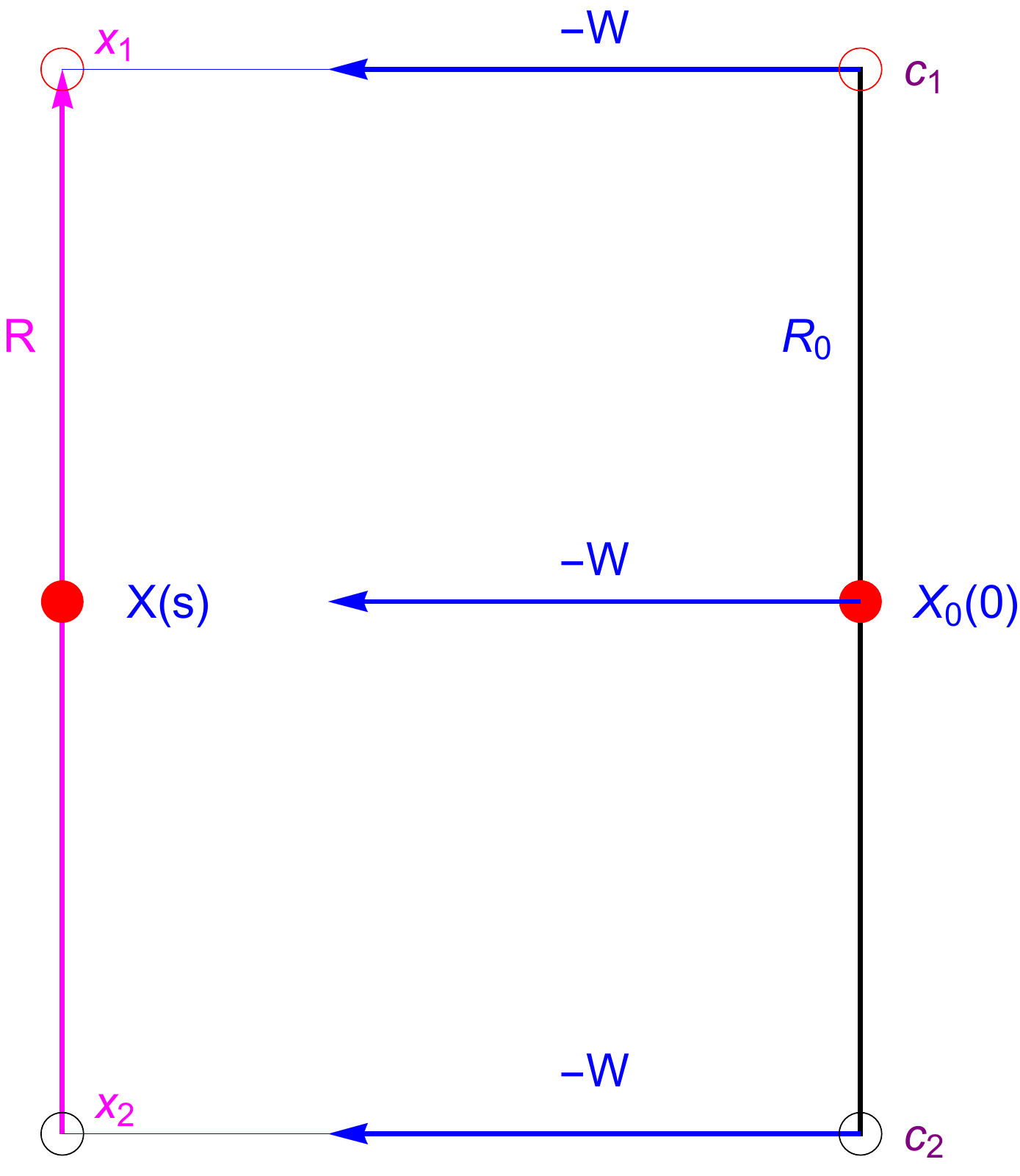}
\\{}
\null\hskip8mm
(a) \hskip55mm (b)\\
\caption{(a)\textit{\small  Two uncharged Carroll particles of equal masses which interact through the momentum-dependent potential $V=V(\bv_a)$ in \eqref{BGLpot} \underline{do move}, namely in opposite directions with constant velocities $\mp\bW^0$. Their CoM $\bX=\bX_0$ is fixed but their relative position, $\bR=\bR(s)$, varies with (Carroll) time.} (b) \textit{\small  For an electric dipole composed of oppositely charged Carroll particles the roles interchanged: the CoM  of the \underline{isolated} system \underline{moves} under the effect of the momentum-dependent \underline{internal} force, whereas their relative coordinate, $\bR=\bR_0$, is rigidly translated.}
\label{2partfig}
}
\end{figure}

In conclusion, while \emph{the center of mass remains fixed  the {relative position}, ${\bR}=\bx_1-\bx_2$ in \eqref{relcoord}
moves with constant velocity $-2\bW^0$}, \eqref{B14grelmot}. Therefore the position coordinates do move as shown in  fig.\ref{2partfig}a, namely with constant but opposite (Carroll) velocities.

The immobility of the CoM is consistent with C-boost symmetry.
Presenting the Hamiltonian action \eqref{GomisHact} as
\beq
\cS=\int\!2\bU\cdot{d\bX}+
\half\bW\cdot {d\bR}
+\half\bW^2ds
\label{GomisHactbis}
\eeq
shows indeed that the infinitesimal Carroll boost
\beq
\delta\bX=0,\quad \delta\bR=0,\quad
\delta\bU = \bbeta,\quad\delta \bW = 0,\quad
\delta s = -\bbeta\cdot{\bX}\,
\label{GomisCboost}
\eeq
changes the Cartan form of \eqref{GomisHact} by a surface term and the conservation of $\bd$, \eqref{CoMmom} with $M=2$, is recovered.
$\bd^{\prime}=0$ implies, conversely, the immobility of the CoM, $\bX$, consistently with \eqref{GomisXR}. No conclusion can be drawn from the conservation of $\bd$, though, about the relative motion: unlike for the CoM, the dynamics of $\bR$ \emph{does} depend on $\bW$: the C-boost symmetry affects the motion of the CoM, but not the relative motion.
\goodbreak

\kikezd{Electric dipoles}.
 The conservation of the center of mass is reminiscent of that of the dipole moment of fractonic systems \cite{Pretkofractons,Seiberg,Gromov:2018nbv,DoshiG,Pena-Benitez:2021ipo,JainJensen,Bidussi} which can easily be incorporated into the multi-Carrollian picture by noticing that the masses $m_a$  can in fact also be viewed  as \emph{charges} which then can take both positive and negative values, $m_a \leadsto q_a$. Then the definition of the center of mass reduces to the familiar expression of the total dipole moment,
\begin{align}
\bd = \sum_{a} q_a\bx_a\,.
\label{NdipoleQ}
\end{align}
 Our general theory then implies that the motion of interacting Carroll particles conserves  the total dipole moment~\eqref{NdipoleQ} provided the interaction is realized via a momentum-dependent two-particle potential~\eqref{Vvpot} -- and then the conservation of \eqref{NdipoleQ} follows from the C-boost symmetry via Noether's theorem.

Let us spell this out  for an electric dipole composed of two opposite electric charges with equal masses, $m_1=m_2=1$ but opposite charges, $q_1=q,\; q_2= -q$. Their total charge is thus zero and no ``center of charge" analogous to the center of mass can be defined.
The Hamiltonian action is the integral of 
\beq 
\cL_{dip}\, {ds} = q(\bv_1 d\bx_1-\bv_2 d\bx_2) - Vds\,,
\label{dipolLag}
\eeq
where the potential is independent of the charges; we have chosen $V = - \dfrac{1}{2}\bW^2$ as before. 
The motions can be determined directly:  the equations of motion are
\beq
\bX^{\prime} =-\frac{\bW}{q}\,,
\qquad
\bR^{\prime} 
= 0\,.
\label{dipoleeqns}
\eeq
Comparison with \eqref{UWXRvareq} then shows that the new casting  amounts simply to \emph{interchanging the CoM and the  relative position}, $\bX \leftrightarrow \half \bR$.
We still have
$\bW=\bW^0=\const$ but the positions now
 \emph{move} with \emph{identical} Carroll velocities,
$
\bx_1^{\prime}= \bx_2^{\prime} = -\bW^0,
$
as shown in fig.\ref{2partfig}b.

Due to the relative sign change the no-motion-for-the-CoM condition \eqref{vNIII} becomes now a ``no-relative-motion condition'', $\bR^{\prime}=0$.
Remarkably, the CoM of an \emph{isolated} dipole is in turn \emph{not fixed}, consistently with the violation of the Kohn condition \cite{Kohn61,KohnI,KohnII,ZHAGK,Inzunza},
\beq
{q_a}/{m_a}
=\const \quad \text{for}\quad a= 1, \dots , N\,,
\label{Kohncond}
\eeq
which is indeed necessary for the center-of-mass decomposition \cite{SSD}. However for a dipole we have instead $q_1/m_1= - q_2/m_2.$

 The dipole is symmetric w.r.t. \eqref{Nboost} which
changes the dipole Lagrangian \eqref{dipolLag} by a surface term ; the associated
 conserved quantity is the \emph{electric dipole momentum}, 
 \beq
\bd_{dip} =q (\bx_1-\bx_2) = q\, \bR\,,
\label{consdipmom}
\eeq
whose  conservation implies $\dot{\bR}=0$, consistently with  the relative position $\bR$ being shifted rigidly with the displacement of the CoM, $\bX(s) = -\bW\,s+\bX_0 $, as shown in fig.\ref{2partfig}b.

\goodbreak

\section{Discussion }\label{Concl}

It has long been believed that particles with Carroll symmetry do not move and present therefore only academic interest.
A rather exotic counter-example with momentum-dependent interaction, \eqref{BGLpot}, is put forward by spelling out the one proposed Bergshoeff et al \cite{Bergshoeff14}. 

Condensed matter people found  that their recently  introduced 
{fractons} are pinned down because of what they call \emph{dipole symmetry} \cite{fractons,Pretkofractons,Seiberg,Gromov:2018nbv,DoshiG,Pena-Benitez:2021ipo}.
 The relation with Carroll was hinted at in \cite{Bidussi}; see also \cite{JainJensen} using a somewhat different terminology.
 These authors identified Carroll symmetry with transformations of time, $t$, which, for a free particle, actually extends to infinite BMS symmetry \cite{BMS,Bagchi,BMSCarr}.
Dipole symmetry acts in turn on the phase of a complex scalar field --- which corresponds to the ``vertical'' action in the Kaluza-Klein-type Bargmann framework \cite{DGH91,DBKP}.

Thus in the free case there is a double Carroll symmetry, along either or the other light-cone coordinate. 
The two can be unified \cite{BMSCarr,HallCarroll} in a Kaluza-Klein type framework \cite{Carrollvs,Carroll4GW}, where a ``vertical'' internal phase variable, $s$, is introduced. Dipole symmetry is indeed identified with $s$-based Carroll symmetry \cite{HallCarroll}.

Interactions break the $t$-based symmetry to  mere  $t$-translations, however the first, linear term of the $s$-based one does survive --- and then yields dipole symmetry.

Returning to immobility, the Carroll symmetry effects only the CoM and implies, for a two-particle Carroll system with the BGL potential \eqref{BGLpot}, the immobility of the latter. The relative coordinates are not affected by the Carroll action \eqref{GomisCboost} and they  can indeed move, as illustrated in sec.\ref{BGLEx}, cf. fig.\ref{2partfig}a.

Replacing the masses by electric charges, the r\^oles of center of mass, $\bX$, and of the relative coordinate, $\bR$, are interchanged; Carroll boosts act on $\bR$ but not on $\bX$, cf. fig.\ref{2partfig}b.

We note in conclusion that immobility is related to Kohn's theorem on decomposing a particle system into center-of-mass and relative coordinates \cite{Kohn61,KohnI,KohnII,ZHAGK,DBKP,Inzunza}, which requires that the charge/mass ratio of all particles be the same,
\eqref{Kohncond}.
Kohn's theorem is rooted in the very structure of the Galilei group \cite{SSD}. In this note we argue, in the Carroll context, in favor of a similar the r\^ole played by C-boosts, \eqref{Cboost}.

At last, we would like to  mention some recent contributions \cite{Figueroa-OFarrill:2023qty,Tadros:2023teq,Kasikci:2023zdn,Ciambelli:2023tzb,Ciambelli:2023xqk,Ecker:2024lur}.

\vskip-3mm

\begin{acknowledgments}\vskip-4mm
The authors are grateful to J-M L\'evy-Leblond for correspondance and to Janos Balog for useful advices. Discussions are acknowledged to Maxim Chernodub. Some of our results were obtained in collaboration with Lo\"\i c Marsot \cite{HallCarroll}. We thank also Alfredo Perez for calling our attention at their recent paper \cite{FPP}. PMZ was partially supported by the National Natural Science Foundation of China (Grant No. 11975320).
\end{acknowledgments}
\goodbreak



\begin{thebibliography}{99}

\bibitem{Leblond}
J. M. L\'evy-Leblond,
``Une nouvelle limite non-relativiste du group de Poincar\'e,''
Ann. Inst. H Poincar\'e {\bf 3} (1965) 1; 

\bibitem{SenGupta}
 V. D. Sen Gupta,
 ``On an Analogue of the Galileo Group,''
 Il Nuovo Cimento {\bf 54} (1966) 512. 
 
\bibitem{BacryLL}
H.~Bacry and J.~Levy-Leblond,
``Possible kinematics,''
J. Math. Phys. \textbf{9} (1968), 1605-1614
doi:10.1063/1.1664490

\bibitem{Inonu}
  E.~In\"on\"u and E.~P.~Wigner,
``On the contraction of groups and their representations,''
  Proc.\ Nat.\ Acad.\ Sci.\  {\bf 39} (1953) 510.

\bibitem{Henneaux79}
M.~Henneaux,
``Geometry of Zero Signature Space-times,''
Bull. Soc. Math. Belg. \textbf{31} (1979), 47-63
PRINT-79-0606 (PRINCETON).

\bibitem{Henneaux82}
M.~Henneaux, M.~Pilati and C.~Teitelboim,
``Explicit Solution for the Zero Signature (Strong Coupling) Limit of the Propagation Amplitude in Quantum Gravity,''
Phys. Lett. B \textbf{110} (1982), 123-128
doi:10.1016/0370-2693(82)91019-X

\bibitem{DGH91}
C.~Duval, G.~W.~Gibbons and P.~Horvathy,
``Celestial mechanics, conformal structures and gravitational waves,''
Phys. Rev. D \textbf{43} (1991), 3907-3922
doi:10.1103/PhysRevD.43.3907
[arXiv:hep-th/0512188 [hep-th]].

\bibitem{Gibbons:2002tv}
G.~Gibbons, K.~Hashimoto and P.~Yi,
``Tachyon condensates, Carrollian contraction of Lorentz group, and fundamental strings,''
JHEP \textbf{09} (2002), 061
doi:10.1088/1126-6708/2002/09/061
[arXiv:hep-th/0209034 [hep-th]].

  
\bibitem{Ancille}
A.~Ngendakumana, J.~Nzotungicimpaye and L.~Todjihounde,
``Group theoretical construction of planar Noncommutative Phase Spaces,''
J. Math. Phys. \textbf{55} (2014), 013508
doi:10.1063/1.4862843
[arXiv:1308.3065 [math-ph]].

\bibitem{MarsotJGP}
L.~Marsot,
``Planar Carrollean dynamics, and the Carroll quantum equation,''
J. Geom. Phys. \textbf{179} (2022), 104574
doi:10.1016/j.geomphys.2022.104574
[arXiv:2110.08489 [math-ph]].

\bibitem{Bergshoeff14}
E.~Bergshoeff, J.~Gomis and G.~Longhi,
\textit{``Dynamics of Carroll Particles,''}
Class. Quant. Grav. \textbf{31} (2014) no.20, 205009
doi:10.1088/0264-9381/31/20/205009
[arXiv:1405.2264 [hep-th]].
See also the preliminary version \cite{GoPa},

\bibitem{GoPa}
J. Gomis and F. Passerini,
``Super Carroll space, Carrollian super-particle and Carrollian super-string'' (2005) (unpublished).

\bibitem{Carrollvs}
  C.~Duval, G.~W.~Gibbons, P.~A.~Horvathy and P.~M.~Zhang,
 ``Carroll versus Newton and Galilei: two dual non-Einsteinian concepts of time,''
Class.\ Quant.\ Grav.\  {\bf 31} (2014) 085016
 doi:10.1088/0264-9381/31/8/085016
 [arXiv:1402.0657 [gr-qc]].
  
\bibitem{Carroll4GW}
C.~Duval, G.~W.~Gibbons, P.~A.~Horvathy and P.~M.~Zhang,
``Carroll symmetry of plane gravitational waves,''
Class. Quant. Grav. \textbf{34} (2017) no.17, 175003
doi:10.1088/1361-6382/aa7f62
[arXiv:1702.08284 [gr-qc]].

\bibitem{Donnay19}
L.~Donnay and C.~Marteau,
``Carrollian Physics at the Black Hole Horizon,''
Class. Quant. Grav. \textbf{36} (2019) no.16, 165002
doi:10.1088/1361-6382/ab2fd5
[arXiv:1903.09654 [hep-th]].

\bibitem{Donnay22}
L.~Donnay, A.~Fiorucci, Y.~Herfray and R.~Ruzziconi,
``Carrollian Perspective on Celestial Holography,''
Phys. Rev. Lett. \textbf{129} (2022) no.7, 071602
doi:10.1103/PhysRevLett.129.071602
[arXiv:2202.04702 [hep-th]].

\bibitem{MarsotHP}
L.~Marsot, P.~M.~Zhang and P.~Horvathy,
``Anyonic spin-Hall effect on the black hole horizon,''
Phys. Rev. D \textbf{106} (2022) no.12, L121503
doi:10.1103/PhysRevD.106.L121503
[arXiv:2207.06302 [gr-qc]].

\bibitem{HallCarroll}
L.~Marsot, P.~M.~Zhang, M.~Chernodub and P.~A.~Horvathy,
``Hall effects in Carroll dynamics,''
Phys. Rept. \textbf{1028} (2023), 1-60
doi:10.1016/j.physrep.2023.07.007
[arXiv:2212.02360 [hep-th]].

\bibitem{fractons}
M.~Pretko,
``Subdimensional Particle Structure of Higher Rank U(1) Spin Liquids,''
Phys. Rev. B \textbf{95} (2017) no.11, 115139
doi:10.1103/PhysRevB.95.115139
[arXiv:1604.05329 [cond-mat.str-el]].

\bibitem{Pretkofractons}
M.~Pretko,
``The Fracton Gauge Principle,''
Phys. Rev. B \textbf{98} (2018) no.11, 115134
doi:10.1103/PhysRevB.98.115134
[arXiv:1807.11479 [cond-mat.str-el]].

\bibitem{Seiberg}
N.~Seiberg,
``Field Theories With a Vector Global Symmetry,''
SciPost Phys. \textbf{8} (2020) no.4, 050
doi:10.21468/SciPostPhys.8.4.050
[arXiv:1909.10544 [cond-mat.str-el]].

\bibitem{Gromov:2018nbv}
A.~Gromov,
``Towards classification of Fracton phases: the multipole algebra,''
Phys. Rev. X \textbf{9} (2019) no.3, 031035
doi:10.1103/PhysRevX.9.031035
[arXiv:1812.05104 [cond-mat.str-el]].
M.~Pretko, X.~Chen and Y.~You,
``Fracton Phases of Matter,''
Int. J. Mod. Phys. A \textbf{35} (2020) no.06, 2030003
doi:10.1142/S0217751X20300033
[arXiv:2001.01722 [cond-mat.str-el]].

\bibitem{DoshiG}
D.~Doshi and A.~Gromov,
``Vortices and Fractons,''
Commun Phys 4, 44 (2021). 
https://doi.org/10.1038/s42005-021-00540-4
[arXiv:2005.03015 [cond-mat.str-el]].

\bibitem{Pena-Benitez:2021ipo}
F.~Pe\~na-Benitez,
``Fractons, Symmetric Gauge Fields and Geometry,''
[arXiv:2107.13884 [cond-mat.str-el]].

\bibitem{Bidussi}
L.~Bidussi, J.~Hartong, E.~Have, J.~Musaeus and S.~Prohazka,
``Fractons, dipole symmetries and curved spacetime,''
SciPost Phys. \textbf{12} (2022) no.6, 205
doi:10.21468/SciPostPhys.12.6.205
[arXiv:2111.03668 [hep-th]].


\bibitem{JainJensen}
A.~Jain and K.~Jensen,
``Fractons in curved space,''
SciPost Phys. \textbf{12} (2022) no.4, 142
doi:10.21468/SciPostPhys.12.4.142
[arXiv:2111.03973 [hep-th]].


\bibitem{FPP}
J.~Figueroa-O'Farrill, A.~P\'erez and S.~Prohazka,
``Carroll/fracton particles and their correspondence,''
JHEP \textbf{06} (2023), 207
doi:10.1007/JHEP06(2023)207
[arXiv:2305.06730 [hep-th]].



\bibitem{Casalbuoni}
R.~Casalbuoni, D.~Dominici and J.~Gomis,
``Two interacting conformal Carroll particles,''
Phys. Rev. D \textbf{108} (2023) no.8, 086005
doi:10.1103/PhysRevD.108.086005
[arXiv:2306.02614 [hep-th]].

\bibitem{Waak} 
B. T. Waak,
``Nuclear systematics and momentum-dependent potential'',
Ph D thesis. Texas Tech University (1972)

 \bibitem{Korinek}
F. Korinek, H. Leeb, M. Braun, S.A. Sofianos, R. M. Adam,
``Momentum dependent nucleon-nucleon potentials via inverse scattering techniques,''
Nucl. Phys. A 607, 123 (1996)
https://doi.org/10.1016/0375-9474(96)00213-8
    
\bibitem{DGGB}
C. B. Das, S. Das Gupta, C. Gale, Bao-An Li, 
``Momentum dependence of symmetry potential in asymmetric nuclear matter for transport model calculations,''
Phys.Rev.C67:034611 (2003)
https://doi.org/10.1103/PhysRevC.67.034611
[arXiv:nucl-th/0212090]

\bibitem{Nara}
Yasushi Nara, Tomoyuki Maruyama, and Horst Stoecker,
``Momentum-dependent potential and collective flows within the relativistic quantum molecular dynamics approach based on relativistic mean-field theory,''
Phys. Rev. C 102, 024913 (2020)
DOI: 10.1103/PhysRevC.102.024913

\bibitem{Csaba}
Cs. S\"uk\"osd (private communication).

\bibitem{SSD}
J.-M. Souriau,
\textsl{Structure des syst\`emes dynamiques}. Dunod (1970, \copyright\,1969);
\textsl{Structure of Dynamical Systems. A Symplectic View of Physics},
translated by C.H.~Cushman-de Vries (R.H.~Cushman and G.M.~Tuynman, Translation Editors), Birkh\"auser, 1997.

\bibitem{Ury}
P.~Horvathy and L.~Ury,
``Analogy between dynamics and statics, related to variational mechanics,''
Acta Physica Acad. Sci. Hung. \textbf{42} (1977), 251-260
doi:10.1007/BF03157493

\bibitem{BMS}
H. Bondi, M. G. van der Burg, and A. W. Metzner,
``Gravitational waves in general relativity. 7.
Waves from axisymmetric isolated systems,''
Proc. Roy. Soc. Lond. A {\bf 269} (1962) 21;
R.~Sachs,
 ``Asymptotic symmetries in gravitational theory,''
 Phys.\ Rev.\  {\bf 128} (1962) 2851.

\bibitem{Bagchi}
  A.~Bagchi,
  ``Correspondence between Asymptotically Flat Spacetimes and Nonrelativistic Conformal Field Theories,''
  Phys.\ Rev.\ Lett.\  {\bf 105} (2010) 171601.
  
 
\bibitem{BMSCarr}
C.~Duval, G.~W.~Gibbons, P.~A.~Horvathy, ``Conformal Carroll groups and BMS symmetry,''
Class. Quant. Grav.  {\bf 31} (2014) 092001
 [arXiv:1402.5894 [gr-qc]].
 


\bibitem{Kohn61}
W.~Kohn,
``Cyclotron Resonance and de Haas-van Alphen Oscillations of an Interacting Electron Gas,''
Phys. Rev. \textbf{123} (1961), 1242-1244
doi:10.1103/PhysRev.123.1242

\bibitem{KohnI}
P.~M.~Zhang and P.~A.~Horvathy,
``Kohn's theorem and Galilean symmetry,''
Phys. Lett. B \textbf{702} (2011), 177-180
doi:10.1016/j.physletb.2011.06.081
[arXiv:1105.4401 [hep-th]].

\bibitem{KohnII}
P.~M.~Zhang and P.~A.~Horvathy,
``Kohn condition and exotic Newton-Hooke symmetry in the non-commutative Landau problem,''
Phys. Lett. B \textbf{706} (2012), 442-446
doi:10.1016/j.physletb.2011.11.035
[arXiv:1111.1595 [hep-th]].

\bibitem{ZHAGK}
P.~M.~Zhang, P.~A.~Horvathy, K.~Andrzejewski, J.~Gonera and P.~Kosinski,
``Newton-Hooke type symmetry of anisotropic oscillators,''
Annals Phys. \textbf{333} (2013), 335-359
doi:10.1016/j.aop.2012.11.018
[arXiv:1207.2875 [hep-th]].


\bibitem{DBKP}
C.~Duval, G.~Burdet, H.~P.~Kunzle and M.~Perrin,
``Bargmann Structures and Newton-cartan Theory,''
Phys. Rev. D \textbf{31} (1985), 1841-1853
doi:10.1103/PhysRevD.31.1841


\bibitem{Inzunza}
L.~Inzunza and M.~S.~Plyushchay,
``Conformal generation of an exotic rotationally invariant harmonic oscillator,''
Phys. Rev. D \textbf{103} (2021) no.10, 106004
doi:10.1103/PhysRevD.103.106004
[arXiv:2103.07752 [quant-ph]].


\bibitem{Ngome}
P.~M.~Zhang, P.~A.~Horvathy and J.~P.~Ngome,
``Non-commutative oscillator with Kepler-type dynamical symmetry,''
Phys. Lett. A \textbf{374} (2010), 4275-4278
doi:10.1016/j.physleta.2010.08.054
[arXiv:1006.1861 [hep-th]].



\bibitem{Figueroa-OFarrill:2023qty}
J.~Figueroa-O'Farrill, A.~P\'erez and S.~Prohazka,
``Quantum Carroll/fracton particles,''
JHEP \textbf{10} (2023), 041
doi:10.1007/JHEP10(2023)041
[arXiv:2307.05674 [hep-th]].


\bibitem{Tadros:2023teq}
P.~Tadros and I.~Kol\'a\v{r},
``Carrollian limit of quadratic gravity,''
Phys. Rev. D \textbf{108} (2023) no.12, 124051
doi:10.1103/PhysRevD.108.124051
[arXiv:2307.13760 [gr-qc]].

\bibitem{Kasikci:2023zdn}
O.~Kasikci, M.~Ozkan, Y.~Pang and U.~Zorba,
``Carrollian supersymmetry and SYK-like models,''
Phys. Rev. D \textbf{110} (2024) no.2, L021702
doi:10.1103/PhysRevD.110.L021702
[arXiv:2311.00039 [hep-th]].


\bibitem{Ciambelli:2023tzb}
L.~Ciambelli and D.~Grumiller,
``Carroll geodesics,''
[arXiv:2311.04112 [hep-th]].


\bibitem{Ciambelli:2023xqk}
L.~Ciambelli,
``Dynamics of Carrollian scalar fields,''
Class. Quant. Grav. \textbf{41} (2024) no.16, 165011
doi:10.1088/1361-6382/ad5bb5
[arXiv:2311.04113 [hep-th]].

\bibitem{Ecker:2024lur}
F.~Ecker, D.~Grumiller, M.~Henneaux and P.~Salgado-Rebolledo,
``Carroll-invariant propagating fields,''
Phys. Rev. D \textbf{110} (2024) no.4, L041901
doi:10.1103/PhysRevD.110.L041901


\end{thebibliography}
\end{document}